\documentclass[aps,prl,twocolumn,groupedaddress,nofootinbib,showpacs]{revtex4}

\usepackage{graphicx}
\usepackage{graphics}
\usepackage{amsfonts}
\usepackage{amsmath}
\usepackage{amssymb}
\usepackage{color}
\usepackage{ulem}
\usepackage{ifthen}

\newcommand{\Qtot}{Q}

\newcommand{\cint}{{\rm C}_{\rm int}}
\newcommand{\cdiff}{{\rm C}_{\rm diff}}

\newcommand{\bfrN}{{\bf r}^N}
\newcommand{\dbfrN}{{\rm d}{\bf r}^N\ }
\newcommand{\bfq}{{\bf q}}
\newcommand{\dbfq}{{\rm d}{\bf q}}
\newcommand{\bfPsi}{\boldsymbol{\Psi}^0}

\newcommand{\mathH}{\mathcal{H}}

\newcommand{\mathU}{\mathcal{U}}
\newcommand{\mathF}{\mathcal{F}}
\newcommand{\mathZ}{\mathcal{Z}}

\newcommand{\CondProb}[2]{P(#1|#2)}

\newcommand{\avg}[1]{\left\langle\displaystyle #1\right\rangle}

\newcommand{\highlightrevision}{false}

\ifthenelse{\equal{\highlightrevision}{true}}
{
\newcommand{\revision}[1]{{\color{red}{#1}}}
\newcommand{\revbar}[1]{{\sout{#1}}}
}
{
\newcommand{\revision}[1]{{\color{black}{#1}}}
\newcommand{\revbar}[1]{}
}

\begin{document}

\title{Charge fluctuations in nano-scale capacitors}

\author{David T. Limmer$^1$}
\author{C\'eline Merlet$^{2,3}$}
\author{Mathieu Salanne$^{2,3}$}
\author{David Chandler$^1$}
\author{Paul A. Madden$^4$}
\author{Ren\'e van Roij$^5$}
\author{Benjamin Rotenberg$^{2,3}$}
\affiliation{$^1$ \small Department of Chemistry, University of California, Berkeley, CA-94720, USA}
\affiliation{$^2$ \small UPMC Univ-Paris06 and CNRS, UMR 7195, PECSA, F-75005, Paris, France}
\affiliation{$^3$ \small R\'eseau sur le Stockage Electrochimique de l'Energie (RS2E), FR CNRS 3459, France}
\affiliation{$^4$ \small Department of Materials, University of Oxford, Parks Road, Oxford OX1 3PH, UK}
\affiliation{$^5$ \small Institute for Theoretical Physics, Utrecht University, 3584 CE Utrecht, The Netherlands }

\date{\today}

\begin{abstract}
The fluctuations of the charge on an electrode  
contain information on the microscopic correlations within the adjacent fluid 
and their effect on the electronic properties of the interface.
We investigate these fluctuations 
using molecular
dynamics simulations in a constant-potential ensemble with histogram reweighting
techniques. \revision{This approach offers in particular an efficient,
accurate and physically insightful route to the differential capacitance 
that is broadly applicable}.
We demonstrate these methods with three  different capacitors:
pure water between platinum electrodes, and a pure as well as a solvent-based 
organic electrolyte each between graphite electrodes. The total charge distributions
with the pure solvent and solvent-based electrolytes
are remarkably Gaussian, while in the pure ionic liquid the total charge
distribution displays distinct non-Gaussian features,
suggesting significant potential-driven changes in the organization of 
the interfacial fluid.
\end{abstract}


\maketitle


The charge of an electrode in contact with a liquid and maintained at a constant
potential undergoes thermal fluctuations that encode information on 
microscopic interfacial processes. Most common applications involving such 
interfaces, such as charge storage in dielectric or electrochemical double layer 
capacitors~\cite{simon2008a}, electrochemistry, water purification, 
or the growing field of ``blue energy''~\cite{brogioli_extracting_2009,
brogioli_prototype_2011,boon_blue_2011} 
utilize only the ability of the metal to aquire an average charge upon
application of voltage. 
However, it is also possible to extract microscopic
information on the interfacial processes from the
fluctuation of the electrode charge,
\revision{both near and far from equilibrium, from the
large-deviation statistics of fluctuations of the electrode charge.}
Our purpose here is to demonstrate this fact and to add to the tools available
to exploit it.
 
As nanoscale devices become widely available, it is essential to better 
understand these fluctuations.
Experimentally, this possibility is \revbar{only}
rarely exploited, with the notable exceptions of electrochemical
noise analysis to infer redox reaction rates and information on corrosion
processes~\cite{bertocci_noise_1995,cottis_interpretation_2001} 
or more recently electrochemical correlation spectroscopy
for single molecule detection and ultralow flow rate measurements in 
nanofluidic channels~\cite{zevenbergen_stochastic_2011,
mathwig_electrical_2012}.
The opportunities offered by such approaches remain however limited by
the theoretical tools to interpret the signal and uncover the underlying 
processes. 

\begin{figure}[h!]
\begin{center}
\vspace{-0.5cm}
\includegraphics[width=7.5cm]{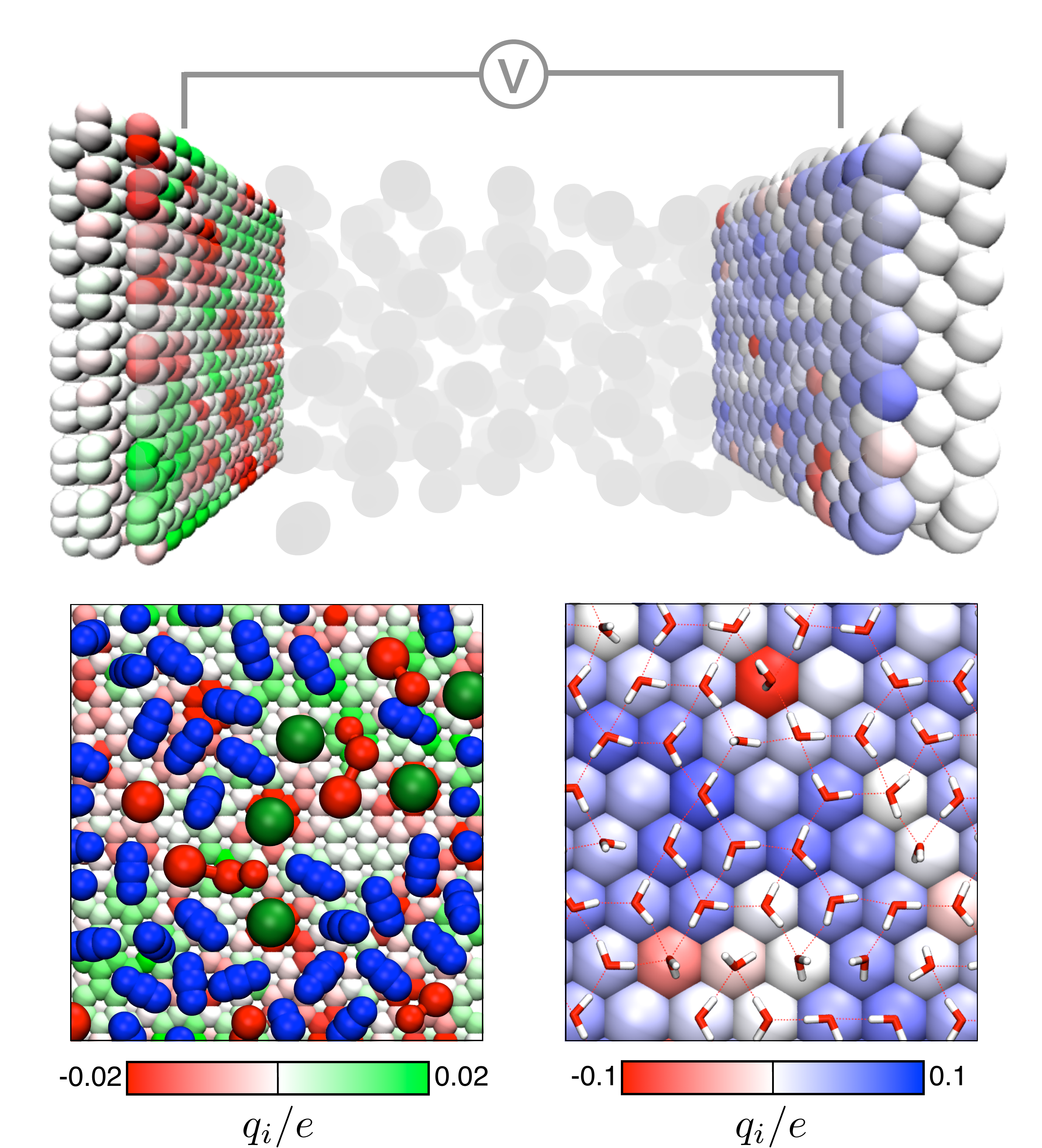}
\vspace{-0.5cm}
\end{center}
\caption{\label{fig:setup}
Each capacitor consists of an electrolyte between two electrodes maintained at a
constant potential difference. The color code on the electrode atoms indicates
the instantaneous charge, $q_i$, with the corresponding scale shown at the
bottom. The left panels show a graphite electrode and at the bottom left is a
representative configuration of the first adlayer of the 1.5~M 
1-butyl-3-methylimidazolium (red) hexafluorophosphate (green) 
in acetonitrile (blue) solution in contact with it. 
The right panels show the 111 crystal facet of a
platinum electrode and at the bottom right is a representative configuration of
the first adlayer of the water in contact with it.
}
\end{figure}

Traditional mean-field treatments~\cite{parsons_review-doublelayer_1990,
kornyshev2007a,lauw_room-temperature_2009,bazant2011a,
skinner_room-temperature_2011} including some models of electric
current fluctuations~\cite{gabrielli_fluctuations_1993,gabrielli_fluctuations_1993-1}, 
ignore the fluctuations we consider.
During the past decade, however, molecular simulations have been
successfully applied to the study of various metallic electrodes
(aluminum, platinum, graphite, nanoporous carbon) and electrolytes (aqueous and
organic solutions,
molten salts, ionic liquids)~\cite{reed2007a,
willard2009a,
vatamanu_molecular_2010,tazi2010a,merlet2011a,merlet2012a}. 
In such simulations, it is essential to account for the polarization of the
electrode by the ions. 
In turn, this polarization screens the (effective) interactions between the ions
and thereby directly affects the structure and dynamics of the 
interface~\cite{merlet2013a}. Analytical models accounting for the image
charge induced on the electrode~\cite{kondrat2011a} remain limited to regular geometries.
Nevertheless, efficient algorithms have been introduced to simulate electrodes
in which the potential is maintained at a constant
value~\cite{siepmann1995a,reed2007a,bonnet2012a}.
The charge on each electrode atom then fluctuates in response to the thermal
motion of the fluid and these fluctuations at any instant are significantly
heterogeneous. See Figure~\ref{fig:setup}.

Let ${\mathH=K({\bf p}^N)+U(\bfrN,\bfq)}$ be 
the microscopic Hamiltonian of the system with ion positions
$\bfrN=\{ {\bf r}_I \}_{I=1\dots N}$, 
ion momenta ${\bf p}^N=\{ {\bf p}_I \}_{I=1\dots N}$ 
and electrode charge distribution $\bfq=\{q_i\}_{i=1\dots 2M}$ with
$2M$ including the atoms of both electrodes. The electrode atoms are fixed in
space. The kinetic part $K$ depends only on the ion momenta and its contribution
to partition functions can be trivially integrated out. Thus in the
following we focus only on the potential part $U$. 
The constant-potential ensemble is defined in terms of the potential of each electrode atom
${\bfPsi=\{\Psi_i^0\}_{i=1\dots 2M}}$.
In this ensemble, the charge distribution $\bfq$ in the electrodes
fluctuates as a result of charge exchange with a reservoir, namely
the external circuit which connects the two electrodes. Charging the capacitor
from $\bfq=0$ to a charge distribution $\bfq$ under fixed $\bfPsi$ 
corresponds to a work exchange $\bfq\cdot\bfPsi$ with this reservoir.
Thus the probability of a state with ion positions $\bfrN$ is
\begin{equation}
\label{eq:PconstPsi-init}
\CondProb{\bfrN}{\bfPsi} = 
\frac{ \int \dbfq\ e^{-\beta U(\bfrN,\bfq) + \beta\bfq\cdot\bfPsi } }
{ \int \dbfrN \dbfq\ e^{-\beta U(\bfrN,\bfq) + \beta\bfq\cdot\bfPsi }  } \; ,
\end{equation}
where $\beta=1/k_{\rm B}T$, with $k_{\rm B}$ Boltzmann's constant and $T$ the temperature.

The integrals can be computed using a saddle point expansion around the charge 
distribution $\bfq^*$ minimizing the term in the exponential, which
satisfies:
\begin{equation}
\label{eq:SetPsi}
\left.\frac{\partial U(\bfrN,\bfq)}{\partial \bfq}\right|_{\bfq=\bfq^*} = \bfPsi \; ,
\end{equation}
{\it i.e.} such that the potential on each atom is the imposed one.
As shown in Supplementary Informations~\cite{SI},
the probability of a state with ion positions $\bfrN$ (and corresponding
charge distribution $\bfq^*$) can be expressed exactly using the
Legendre transform $\mathU(\bfrN,\bfPsi)=U(\bfrN,\bfq^*)-\bfq^*\cdot\bfPsi$ as 
$\CondProb{\bfrN}{\bfPsi}=e^{-\beta \mathU(\bfrN,\bfPsi) } / \mathZ(\bfPsi)$,
with 
{$\mathZ(\bfPsi) = \int \dbfrN e^{-\beta \mathU(\bfrN,\bfPsi)}$}.

In practice one is only interested in the case
where the potential can take only two values, namely $\Psi_i^0=\Psi_+^0$ for all atoms
in the positive electrode and $\Psi_i^0=\Psi_-^0$ for all atoms in the negative
electrode. This corresponds to the condition of a constant potential inside a
metal (perfect conductor).
In that case the additional energy term simplifies to 
$\bfq^*\cdot\bfPsi = \sum_{i\in\pm} q_i^*\Psi_i^0 = \Qtot\Delta\Psi$ ,
with $\Qtot=\Qtot^+ =-\Qtot^-$ the total charge of the
positive electrode and $\Delta\Psi=\Psi_+^0-\Psi_-^0$ the potential difference 
between the electrodes. Note that the sign convention to label the electrodes does
not matter. 
Moreover, the probability of a state, hence any
observable property, depends only on $\Delta\Psi$ and not on the absolute value
of the potentials, which are defined with respect to a reference electrode not
present in the system and which provides charge to the 
electrodes. Using the above result, we finally rewrite the probability as
\begin{equation}
\label{eq:PconstPsi}
\CondProb{\bfrN}{\Delta\Psi} = 
\frac{ e^{-\beta U(\bfrN,\bfq^*) + \beta\Qtot\Delta\Psi} }{ \mathZ(\Delta\Psi) } \; ,
\end{equation}
with the partition function
\begin{equation}
\label{eq:ZconstPsi}
\mathZ(\Delta\Psi) 
= e^{-\beta \mathF(\Delta\Psi)}
= \int \dbfrN e^{-\beta U(\bfrN,\bfq^*) + \beta\Qtot\Delta\Psi} \; ,
\end{equation}
and $\mathF$ the associated free energy.
In this ensemble, the average value of any observable $A(\bfrN,\bfq^*)$ is
computed as
\begin{equation}
\label{eq:avgA}
\avg{A} = \int \dbfrN \CondProb{\bfrN}{\Delta\Psi} A(\bfrN,\bfq^*) \; ,
\end{equation}
where one should keep in mind that the charge distribution $\bfq^*$
is not a free variable, as it is determined for each ion configuration $\bfrN$
by Eq.~\ref{eq:SetPsi}. The average total charge determines the
integral capacitance ${\cint = \avg{\Qtot}/\Delta\Psi}$,
whereas the differential capacitance is related to the variance
of the total charge distribution:
\begin{align}
\label{eq:capa-fluct}
\cdiff &= \frac{\partial \avg{\Qtot}}{\partial \Delta\Psi }
= \beta\avg{\delta\Qtot^2} \; ,
\end{align}
with $\delta\Qtot=\Qtot-\avg{\Qtot}$.
This fluctuation-dissipation relation, 
which can be derived by considering the derivatives of $\mathZ$
with respect to $\Delta\Psi$~\cite{SI}, is known in electronics 
as the Johnson-Nyquist relation~\cite{johnson1928a,nyquist1928a}. 
\revbar{In analogy with the connection between the compressibility of a system
and the small wave-vector limit of the structure factor, 
we}
\revision{We} can also show that the capacitance is related to the
charge-charge structure factor inside the electrode~\cite{SI}:
\begin{align}
\lim_{k\to0}S_{qq}(k) 
&= \frac{\cdiff k_{\rm B} T}{M\avg{\delta q^2}} \;,
\end{align} 
with $\avg{\delta q^2}=\avg{q^2}-\avg{q}^2$ 
the variance of the distribution of the charge per atom.
This result holds for both electrodes, \revbar{(with the same $\cdiff$),} 
even though $S_{qq}(k)$ may differ for non-zero wave-vectors. 
\revbar{as the adsorbed fluid is free to adopt different structures on the two
electrodes.}

The algorithm we use to simulate a metallic electrode maintained at a constant
potential follows from the work of Siepmann and Sprik~\cite{siepmann1995a},
later adapted by Reed {\it et al.} to the case of electrochemical 
cells~\cite{reed2007a}. 
The electrode consists of explicit
atoms bearing a Gaussian charge distribution 
$\rho_i({\bf r})=q_i^* \eta^3 \pi^{3/2} \exp
\left(-\mid {\bf r}-{\bf r}_i\mid^2 \eta^2 \right)$,
where $\eta^{-1}$ is the width of the distribution and where the
atomic charge $q_i^*$ of each atom is determined at each time step of the 
simulation  by minimizing
$U_c - \sum_{i\in\pm}  q_i\Psi^0_i$, 
with $U_c$ the Coulomb energy,
with respect to all the variable charges simultaneously. 
Forces acting on the ions are then computed using the minimizing charges. 

The distribution of the total charge $\Qtot$ in the constant-potential ensemble is:
\begin{align}
\label{eq:PofQtot}
\CondProb{\Qtot}{\Delta\Psi} 
&= \int \dbfrN \CondProb{\bfrN}{\Delta\Psi} \delta\left( \Qtot -\sum_{i\in+}q_i \right) 
\end{align}
with $\delta$ the Dirac distribution. 
The distributions of the total charge can be sampled directly from simulations
at the corresponding potentials. However, this sampling is limited to values of
the total charge \revision{that} are close to the average $\avg{\Qtot}$. 
A more accurate estimate can be obtained by combining the data from the 
simulations performed for various potential differences using histogram reweighting.
Indeed, one can show that
\begin{align}
-\ln\CondProb{\Qtot}{0} &= -\ln\CondProb{\Qtot}{\Delta\Psi}
+ \beta\Qtot\Delta\Psi + \beta\Delta\mathF \;,
\end{align}
with $\Delta\mathF=\mathF(\Delta\Psi) - \mathF(0)$
the difference in free energy (defined by Eq.~\ref{eq:ZconstPsi}).
Each simulation under an applied potential thus provides
an estimate of the charge distribution at any other potential,
up to the unknown constants $\mathF(\Delta\Psi)$,
which are determined self-consistently in the weighted histogram analysis
method (WHAM)~\cite{ferrenberg_optimized_1989,kumar_weighted_1992,roux_calculation_1995}.
Such an approach is well established in other contexts, \revbar{such as simulations
performed at different temperatures,} but has not yet been considered for
simulations in the constant-potential ensemble.

We investigate several capacitors illustrated in Figure~\ref{fig:setup}:
pure water between platinum electrodes and 
an organic electrolyte, 
1-butyl-3-methylimidazolium hexafluorophosphate (BMI-PF$_6$),
either as a pure ionic liquid or as
a 1.5~M  solution in acetonitrile (MeCN), between graphite electrodes.
Details on the systems and molecular models can be found in the
Supplementary Information~\cite{SI}.
These combinations of electrodes and electrolytes offer a large contrast of properties:
The former is a dielectric capacitor containing only neutral molecules, 
while the latter contain ions in the gap and are hence ``double-layer'' capacitors.
In addition, in the former case the water molecules form hydrogen-bonds
and have a size comparable to that of the electrode atoms, 
while in the latter all ions and molecules are large so that 
the electrode appears rather smooth on their scale.
Figure~\ref{fig:setup} also shows the local charge distribution 
on one of the electrodes for instantaneous configurations of the solvent-based 
systems under a potential difference.
It is strikingly heterogeneous and strongly correlated with the local structure
of the adsorbed fluid.  

\begin{figure}[h!]
\begin{center}
\includegraphics[width=8cm]{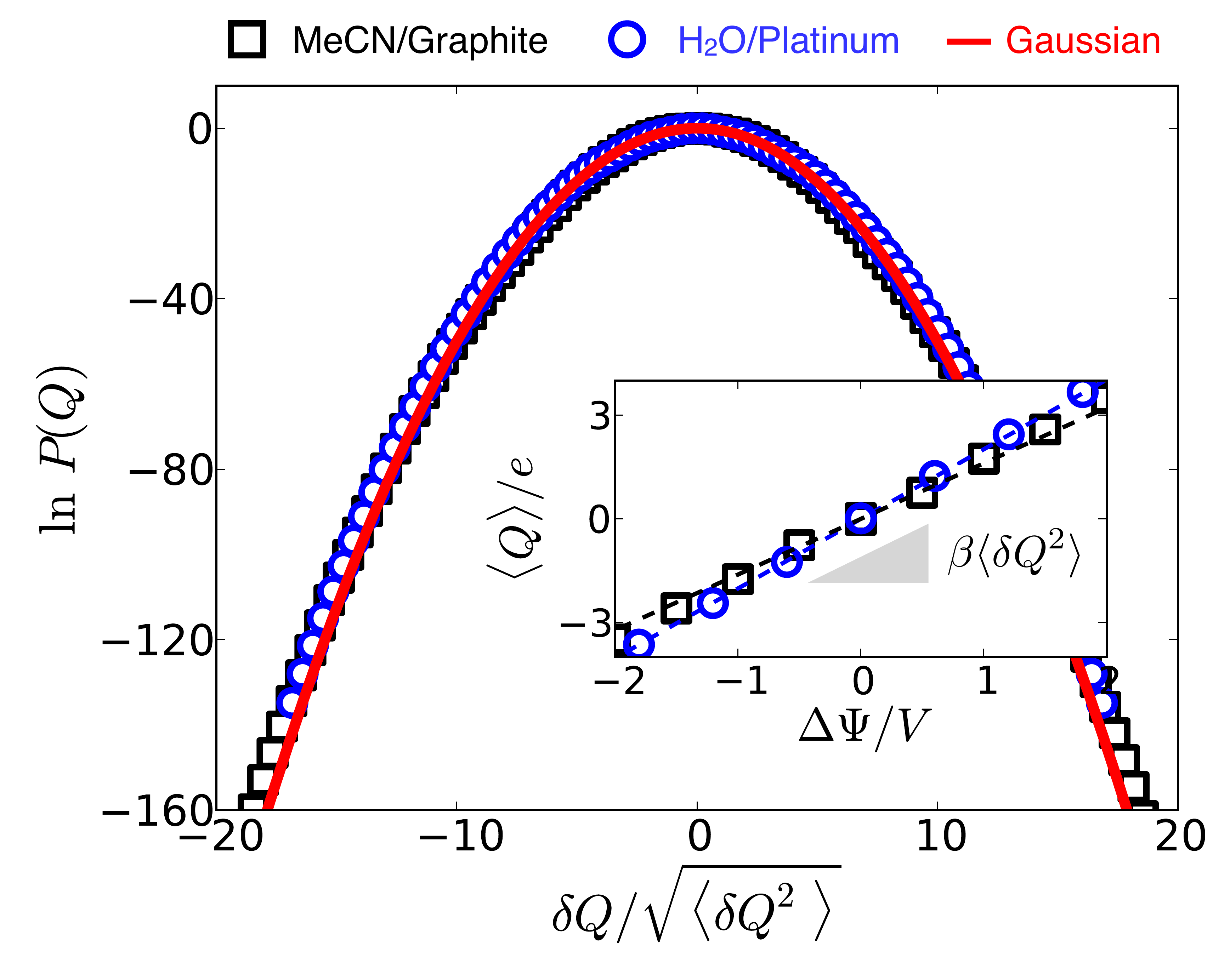}
\vspace{-1cm}
\end{center}
\caption{\label{fig:PofQ}
Probability distribution of the total charge $\Qtot$. 
on the electrodes at $\Delta\Psi=0$~V
for the \revision{acetonitrile (MeCN) electrolyte and water-based capacitors.}
The data is reported as a function of $\delta\Qtot/\sqrt{\avg{\delta\Qtot^2}}$.
\revbar{,
with $\delta\Qtot=\Qtot-\avg{\Qtot}$ and where the variance is
40\% larger in the H$_2$O/Pt case.}
The red line is a Gaussian distribution with the same mean and variance.
\revbar{In both cases, the distribution is Gaussian.}
The inset compares the average charge as a function of voltage
from simulations (symbols) with lines of slope
$\beta\avg{\delta Q^2}$: This illustrates the linear response of
both systems and the validity of the fluctuation-dissipation
relation Eq.~(\ref{eq:capa-fluct}).
}
\end{figure}

Figure~\ref{fig:PofQ} shows that fluctuations of the total charge on the
electrodes for both solvent-based systems are Gaussian to a remarkable degree.
These statistics imply the validity of the linear response theory over the range
of charges shown in Figure~\ref{fig:PofQ}.
The inset shows that $\avg{\Qtot}$ is indeed proportional to
the applied potential $\Delta\Psi$ with a slope $\beta\avg{\delta\Qtot^2}$.
Such a comparison not only provides information on the physical properties of
these two capacitors, but also demonstrates the
relevance of this new approach to determine the differential capacitance.
The latter is $\approx$40\% larger in the water/Pt case (3.2 vs 2.3 and 2.1~$\mu$F.cm$^{-2}$
for the graphite capacitors with the solution in MeCN and pure ionic liquid,
respectively). 
Continuum theory for water between electrodes in the simulated geometry
(distance $d=5.2$~nm between the surfaces), using the permittivity of the 
SPC/E water model, predicts a capacitance 
$\epsilon_0\epsilon_r/d=11.4$~$\mu$F.cm$^{-2}$,
indicating that the molecular nature of the interface\revision{, which
suppresses dipole fluctuations on the surface~\cite{limmer_hydration_2013},} 
plays an important role 
in the overall capacitance (the effective permittivity in the bulk region
agrees well with that of SPC/E~\cite{willard2009a}).

\begin{figure}[h!]
\begin{center}
\includegraphics[width=8cm]{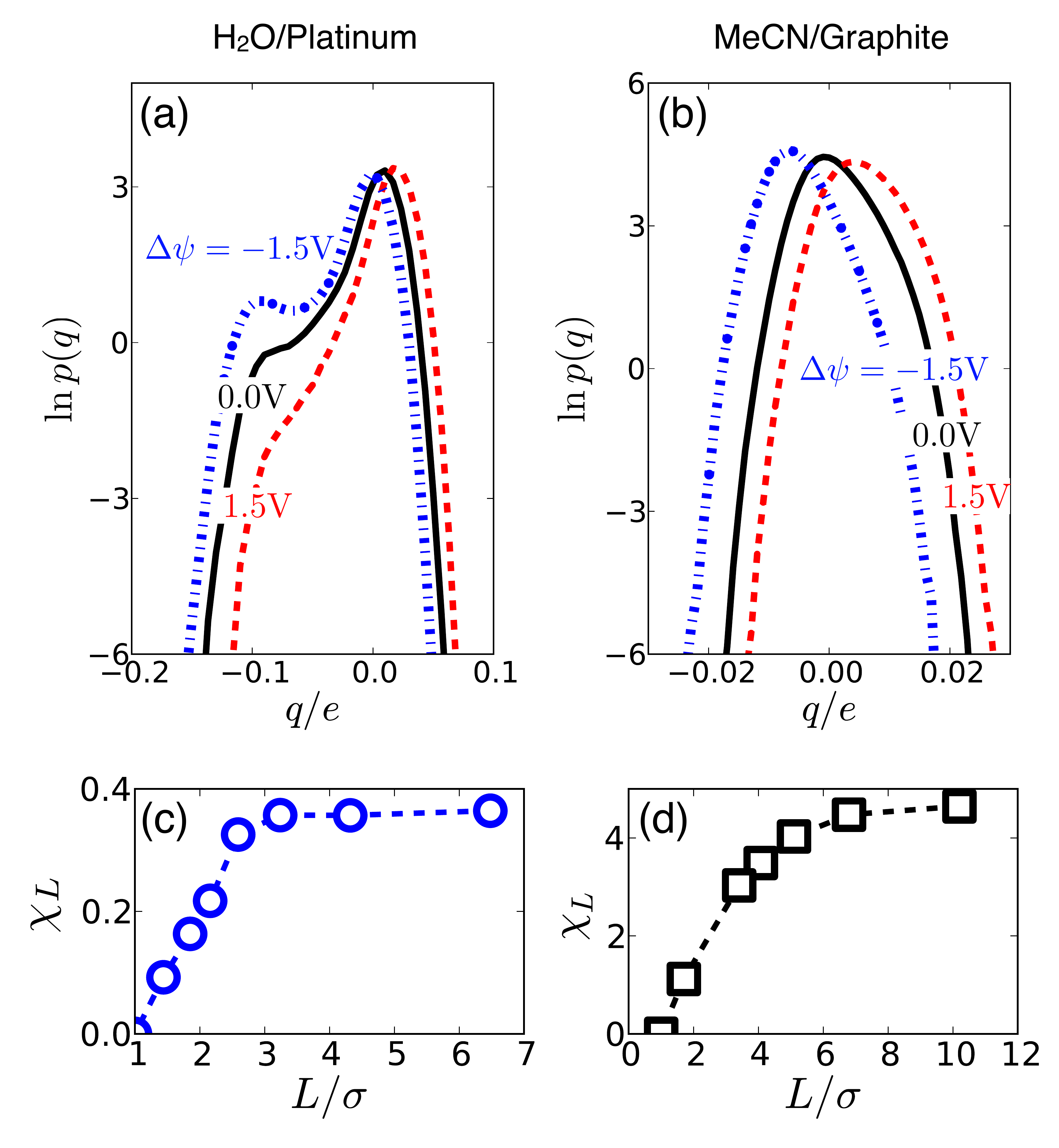}
\vspace{-1cm}
\end{center}
\caption{\label{fig:PofQatom}
Distribution of the charge, $q_i$, on electrode atoms for the H$_2$O/platinum (a)
and MeCN-based organic electrolyte/graphite (b) capacitors,
inside the electrodes in the absence and presence
of voltage ($\Delta\Psi=\pm1.5$~V refer to 
the positive and negative electrodes for 1.5~V). 
(c) and (d)
Variation of the charge fluctuations, $\chi_L$, with increasing electrode area
in units of the electrode atom diameter $\sigma$ (see text). 
}
\end{figure}

The Gaussian behaviour suggests that the charging process for both systems 
arises from \revision{microscopic events that
are correlated over only small lengthscales, those comparable to sizes of molecules.} 
The local charge induced on the
electrode by an interfacial molecule or ion can be analyzed in term of the
distribution of individual charges of the 
electrode atoms. These distributions
are reported for both systems as a function of potential in
Figure~\ref{fig:PofQatom}. The bimodal distribution in the case of water at Pt
arise from the two possible orientations of OH bonds with respect
to the surface, which are asymmetric between the positive and negative electrodes
and evolves with the potential, as the macroscopic electric field favors or
hinders the formation of a hydrogen bond with the surface~\cite{willard2009a}.
For the organic electrolyte on graphite the behaviour is not bimodal, but
the distributions are not Gaussian either, where the non-Gaussianity
stems from the shape- and size asymmetry
of the ions as well as from the dipolar charge distribution of
the acetonitrile molecule. As expected, the larger local charges are induced by
nearby ions rather than solvent molecules. 
As the potential changes,
the main change in the distribution is a shift of its mean, rather than its
shape, as a result of the gradual change in local composition of the 
interfacial fluid. 

The crossover from the non-Gaussian behaviour of the local charge to the
Gaussian distribution of $\Qtot$ suggests the existence of a correlation length
for the charge distribution inside the electrode, which can be determined by
analyzing
\begin{align}
\label{eq:chiL}
\chi_L &= \frac{\avg{\delta Q^2}_L}{\avg{\delta q^2}}\frac{\sigma^2}{L^2}-1
 \; ,
\end{align}
where $\avg{\cdot}_L$ is an average over a piece of the electrode $L\times L$ in
area, the equivalent electrode atom diameter $\sigma=\sqrt{A/M}$ 
with $A$ the electrode area and $M$ the corresponding number of atoms.
\revbar{,
and where as above $\avg{\delta q^2}$ and $\avg{\delta Q^2}$ are 
the one-body and total charge fluctuations, respectively.}
For large enough observation area, the distribution is Gaussian with a variance
proportional to the area, as expected from the extensivity of the capacitance.
The correlation lengths amount to 2-3 water molecules on Pt, consistent with
the surface hydrogen bond network 
(see Figure~\ref{fig:setup})~\cite{willard2009a,limmer_hydration_2013}, and 
$\approx6$ carbon atoms, 
\revision{consistent with the size of the ions.}

\begin{figure}[h!]
\begin{center}
\vspace{-0.2cm}
\includegraphics[width=8.5cm]{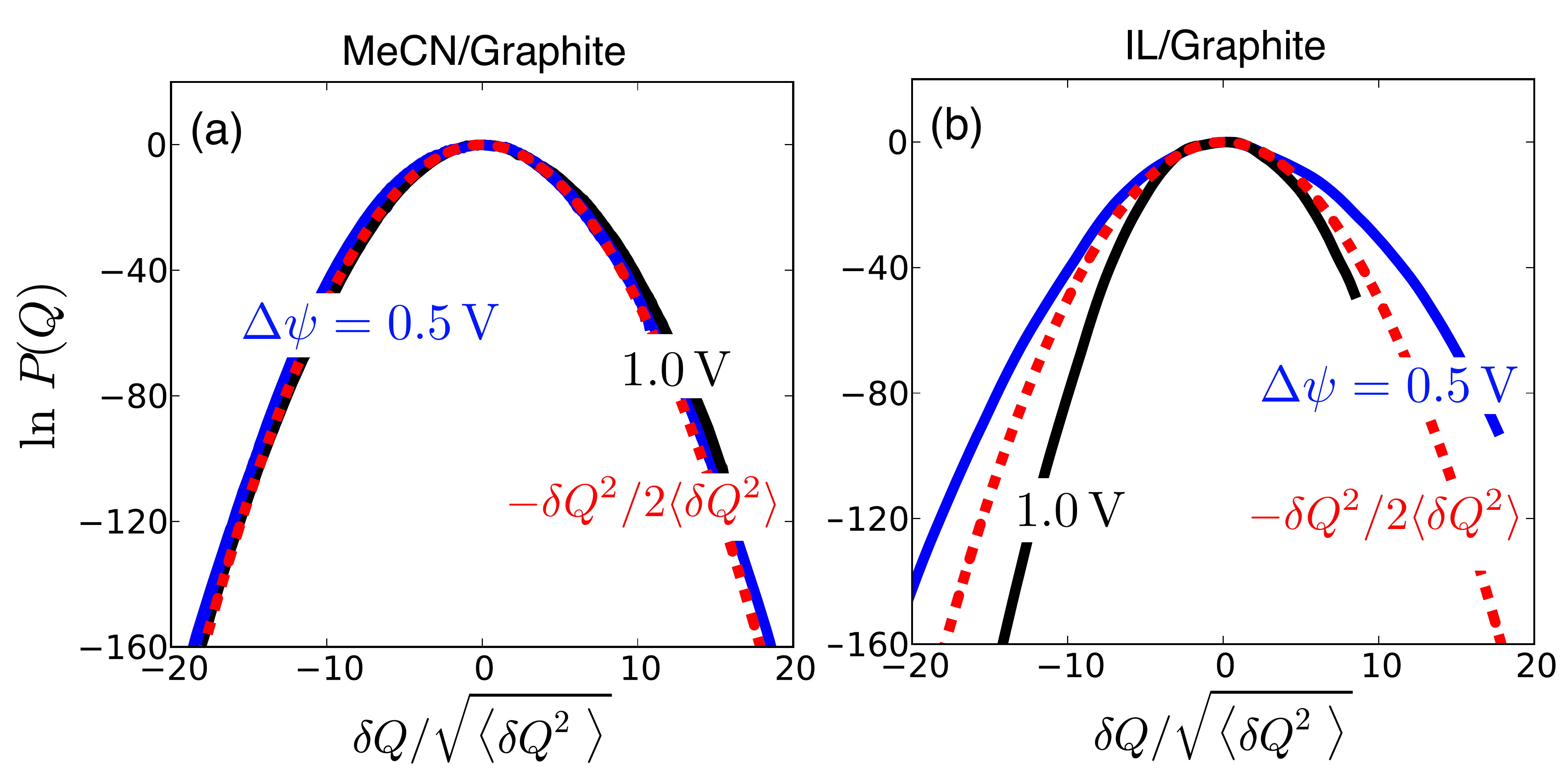}
\vspace{-0.7cm}
\end{center}
\caption{\label{fig:purevsMeCN}
Distribution of the total charge 
for the graphite capacitors, with the MeCN-based electrolyte (a) 
and pure ionic liquid (b).
The results for two applied potentials are compared with
Gaussian distributions with the same variance.
Note that the variance is the same for both potentials in (a)
but is larger at 1~V than at 0.5~V in (b).
}
\end{figure}

The distribution of the total charge is not always Gaussian.
Figure~\ref{fig:purevsMeCN} compares the distributions at $\Delta\Psi=0.5$ 
and 1~V for graphite capacitors with the MeCN-based electrolyte 
and the pure ionic liquid.
While in the former case the distribution is Gaussian with the same
variance for both voltages, 
\revision{for} the pure ionic liquid this variance increases by a factor of about 2.3
between 0.5 and 1~V. These large fluctuations are reflective of correlations
between ions that are not present at low concentration. While the nature of
these correlations is beyond \revbar{of} the scope of this work, we note
that correlations exist that span the electrode sizes we consider here and cause
the fluctuations of the total charge on the electrode to be more or less
probable than if it was determined from the sum of many uncorrelated charge
centers. These aspects of the pure ionic liquid will be considered in detail
elsewhere~\cite{inprep}.

\revision{This sheds new light on the role of correlations
on the differential capacitance~\cite{fedorov_towards_2008,
lamperski_electric_2009,skinner_capacitance_2010,druschler_new_2012}:
It measures the susceptibility for the
interfacial layer of fluid (at a given potential) to develop charge in response
to an increase in the applied potential. It will therefore be low if the fluid
is in a particularly stable configuration and resistant to reorganisation and
large if the fluid has the capability to make a substantial redistribution of
charge over the interfacial region.
}

Combining simulation in the constant-potential ensemble
with histogram reweighting techniques has allowed to investigate correlations in
the adsorbed fluid and their influence on the electronic properties of the
interface. It further provides a unique way to determine the differential 
capacitance, more accurately than previously, and
from a simulation at a {\it single} value of $\Delta\Psi$.
\revision{These methods are based on general principles of statistical
mechanics and as we have shown are widely applicable~\cite{SI}.}
This might prove useful for the study of complex systems such as
nanoporous carbon electrodes where the charging mechanism differs
from the planar graphite case investigated here~\cite{merlet2012a}.
\revision{Generalization of this approach
to dynamic properties~\cite{nieto-draghi_histogram_2005} may lead to new insight on
frequency-dependent capacitance measurements and 
voltage-dependence of lubricating properties of IL films on
metals~\cite{sweeney_control_2012}. }



\section*{Acknowledgements}

BR and DC acknowledge financial support from the France-Berkeley Fund
under grant 2012-0007.
CM, MS and BR acknowledge financial support from the French Agence Nationale de
la Recherche (ANR) under grant ANR-2010-BLAN-0933-02. 
We are grateful for the computing resources on JADE (CINES, French National HPC)
obtained through the project x2012096728.
DTL was supported by the Helios Solar Energy Research Center, which is supported
by the Director, Office of Science, Office of Basic Energy Sciences of the U.S.
Department of Energy under Contract No. DE-AC02-05CH11231.


\newpage
\thispagestyle{empty}
\mbox{}
\newpage

\onecolumngrid

\appendix

\renewcommand{\theequation}{S.\arabic{equation}}
  \setcounter{equation}{0}  

\renewcommand{\figurename}{Supplementary Figure}

\renewcommand{\tablename}{Supplementary Table}

\section*{\Large Supplementary Material}

\begin{center}
\textbf{
\Large Charge fluctuations in nano-scale capacitors }

\vspace*{.5cm}
David T. Limmer,
C\'eline Merlet,
Mathieu Salanne,
David Chandler, \\
Paul A. Madden,
Ren\'e van Roij and
Benjamin Rotenberg
\end{center}


\subsection*{\large Simulation details}

\subsubsection*{\large Dielectric capacitor}

Parameters for the water on platinum system are the same as in our previous
studies~\cite{limmer_hydration_2013,willard_characterizing_2013}. 
In this model the water water interactions are described by the
SPC/E water model~\cite{berendsen_missing_1987}
and those between water and the metal atoms are described
by the Siepmann and Sprik potential~\cite{siepmann1995a}. 
Both electrodes are modeled as three
layers of an FCC crystal with the 111 face in contact with the aqueous solution,
consisting of 1008 atoms with nearly 1600 water molecules. The lattice constant
is 3.92~\AA\ and the total system size is $3.2\times3.3\times5.3$~nm$^3$. The system is
periodically replicated in the $x$ and $y$ directions.

Molecular dynamics simulations were conducted in the NVT ensemble using a time
step of 2~fs and a Nose-Hoover thermostat with a time constant of 5~ps and a
temperature of 298~K. The system is initially equilibrated at constant pressure.
Values of the potential used are (0.0,Ê0.195,Ê0.39,Ê0.585, 0.78,Ê
0.975,Ê 1.17,Ê 1.365, 1.56,Ê 1.755,Ê 1.95, 2.145) V. Simulations are
equilibrated at each target potential and then run for 5~ns.

\subsubsection*{\large Double-layer capacitors}

Molecular dynamics simulations are conducted on two different electrolytes
surrounded by model graphite electrodes: pure BMI-PF$_6$
and its corresponding 1.5~M solutions with acetonitrile (MeCN) as a solvent. 
All molecules are represented by a coarse-grained model in which the forces are 
calculated as the sum of site-site Lennard-Jones potential and coulombic interactions. 
Parameters for the ions and carbon atoms are the same as in our previous
works~\cite{merlet2011a,merlet2012a,merlet2012b,merlet2013b}.
In this model, developed by Roy and Maroncelli~\cite{roy2010a},
three sites are used to describe the MeCN and the cation, while the anions are treated 
as spheres. The model for MeCN was developed by Edwards \emph{et al.}~\cite{edwards1984a}.
The parameters are summarized in Supplementary Table~\ref{tab:param}.
Each electrode is modelled as three fixed graphene layers, with a distance
between carbon atoms within each layer of 1.43~\AA\ and a distance between
layers of 3.38~\AA. The electrolyte is
enclosed between two planar electrodes and two-dimensional periodic boundary
conditions are applied, {\it i.e.} there is no periodicity in the direction
perpendicular to the electrodes. 

\vspace{0.5cm}
\begin{table*}[!ht]
\begin{center}
\begin{tabular}{|c|c|c|c|c|c|c|c|}
\hline
Site                            & C1     & C2     & C3     & PF$_6^-$ & N      & C     & Me \\
\hline
q (e)                           & 0.4374 & 0.1578 & 0.1848 & -0.78    & -0.398 & 0.129 & 0.269 \\
\hline
M (g.mol$^{-1}$)                & 67.07  & 15.04  & 57.12  & 144.96   & 14.01  & 12.01 & 15.04 \\
\hline
$\sigma_i$ (\r{A})              & 4.38   & 3.41   & 5.04   & 5.06     & 3.30   & 3.40  & 3.60 \\  
\hline
$\varepsilon_i$ (kJ.mol$^{-1}$) & 2.56   & 0.36   & 1.83   & 4.71     & 0.42   & 0.42  & 1.59 \\
\hline
\end{tabular}
\end{center}
\caption{Force-field parameters for the molecules of the
electrolytes~\cite{merlet2011a,edwards1984a,roy2010a} (geometries of the
molecules are available in the aforementioned publications). 
C1, C2 and C3 are the three sites of the BMIM$^+$ cation, while Me is the methyl
group of acetonitrile.
Site-site interaction energies are given by the sum of a Lennard-Jones 
potential and coulombic interactions
$u_{ij}(r_{ij})=4\varepsilon_{ij}[(\frac{\sigma_{ij}}{r_{ij}})^{12}-(\frac{\sigma_{ij}}{r_{ij}})^6]+\frac{q_iq_j}{4\pi\varepsilon_0r_{ij}}$
where $r_{ij}$ is the distance between sites, $\varepsilon_0$ is the
permittivity of free space and crossed parameters are calculated by
Lorentz-Berthelot mixing rules. The parameters for the carbon atoms of the
graphite electrodes are $\sigma_{\rm C}$ = 3.37~\r{A} and $\varepsilon_{\rm C}$
= 0.23~kJ.mol$^{-1}$~\cite{cole1983a}.}
\label{tab:param}
\end{table*}

\newpage
Molecular dynamics simulations were conducted in the NVT ensemble
using a time step of 2~fs and a Nos\'e-Hoover thermostat~\cite{martyna1992a}
with a time constant of 10~ps. The Ewald summation is done consistently with the
two-dimensional periodic boundary conditions~\cite{reed2007a,gingrich2010a}.
Pure ILs and electrolyte solutions are simulated at 400~K and 298~K,
respectively. Table~\ref{cells} gathers the lengths and number of molecules 
for the simulation cells.
The algorithm used to maintain the potential constant is described in the main text. 
Five values of potential differences were considered for the MeCN
based electrolyte ($\Delta\Psi=0.0$, 0.5, 1.0, 1.5 and 2.0~V). 
Ten values were simulated for the pure ionic liquid 
($\Delta\Psi=0.0$, 0.2, 0.5, 0.75, 1.0, 1.25, 1.5, 1.75, 1.85 and 2.0~V) in
order to ensure a good overlap between the histograms for $\Qtot$, as required
for the histogram reweighting. 
For each simulation, a 200~ps equilibration is followed by a 5~ns production run
for the pure ionic liquid (1~ns for the MeCN based electrolyte for non-zero
voltages) from which configurations are sampled every 0.2~ps.

\vspace{0.5cm}
\begin{table*}[!ht]
\begin{center}
\begin{tabular}{|c|c|c|c|c|}
\hline
Electrolyte & Temperature (K) & N$_{\rm ions}$ & N$_{\rm MeCN}$ & $L_z$ (nm) \\
\hline
[BMI][PF$_6$] & 400 & 320 & --- & 12.32 \\
\hline
MeCN-[BMI][PF$_6$] & 298 & 96 & 896 & 12.27 \\
\hline
\end{tabular}
\end{center}
\caption{Simulation temperature, number of ion pairs, number of MeCN molecules
and lengths of the simulation cell in the direction perpendicular to the
graphite electrodes for the two electrolytes. The lengths in the $x$ and $y$
directions are the same for all the cells and are equal to 3.22~nm and 3.44~nm
respectively.}
\label{cells}
\end{table*} 

{\revision{
\subsubsection*{\large {System size effects}}

Perpendicular to the plane of the electrode, we have ensured that
in all systems the (number and charge) densities and Poisson potential are uncorrelated
from the wall: They decay to their bulk isotropic values in the center of
the capacitor.
For both solvent-based systems, the system sizes we consider are larger than the
relevant correlation lengths, which are small, as computed in Fig~3. Therefore
it is expected that no significant finite size effects exist in the plane of
the electrode. For the pure ionic liquid, a 3.97~nm by 4.30~nm system size has been
simulated and the capacitance computed at $\Delta\psi=1.0$~V was found
to be in agreement with that computed with the system detailed in the text.

}}

\subsection*{\large Derivation of the fluctuation-dissipation relation}

The average charge is related to the derivative of the partition
function $\mathZ(\Delta\Psi)$ defined by Eq.~(4) of the main text:
\begin{equation}
\label{eq:1stDeriv}
\avg{\Qtot} = \frac{1}{\mathZ} \int \dbfrN 
e^{-\beta U(\bfrN,\bfq) + \beta\Qtot\Delta\Psi} \Qtot
= k_BT \frac{1}{\mathZ} \frac{\partial\mathZ}{\partial \Delta\Psi} \; ,
\end{equation}
while the average square charge $\avg{\Qtot^2}$ is related to
its second order derivative:
\begin{equation}
\label{eq:2ndDeriv}
\avg{\Qtot^2} = (k_BT)^2 
\frac{1}{\mathZ}\frac{\partial^2\mathZ}{\partial \Delta\Psi^2} \; .
\end{equation}
The differential capacitance is defined as
\begin{equation}
\label{eq:Cdiff}
\cdiff = \frac{\partial\avg{\Qtot}}{\partial \Delta\Psi} \; .
\end{equation}
Taking the derivative of Eq.~\ref{eq:1stDeriv} with respect to $\Delta\Psi$
and using Eq.~\ref{eq:2ndDeriv}, one finds after elementary algebra that:
\begin{equation}
\label{eq:capa-fluct}
\avg{\Qtot^2} - \avg{\Qtot}^2 = \cdiff \times k_BT \; ,
\end{equation}
which is the fluctuation-dissipation relation~(6) of the main text.

\subsection*{\large Charge-charge structure factor inside the electrode}


The charge distribution inside the electrodes is quantified
by the charge-charge structure factor 
\begin{align}
 S_{qq}({\bf k})  &= 
\frac{1}{{\revision{M}}\avg{(\delta q)^2}}
 \avg{ \left| \sum_{l}\delta q_l\  e^{-i{\bf k}\cdot{\bf r}_{l}} \right|^2 } \,
,
\end{align}
where the sum runs over electrode atoms $l$, 
$\delta q_{l}=q_{l} - \avg{q}$ with
$q_{l}$ their charge and ${\bf r}_{l}$ their position.
The small wave-vector limit of $S_{qq}({\bf k})$ 
is related to the capacitance of the system. 
Indeed, consider the variance of the total charge:
\begin{align}
\avg{\Qtot^2}-\avg{\Qtot}^2 &= 
\avg{\sum_{l,m} (\avg{q} + \delta q_l ) ( \avg{q} + \delta q_m ) } 
- \avg{ \sum_{l} (\avg{q} + \delta q_l )}^2
\nonumber \\
&=\avg{ \sum_{l,m}\delta q_l \delta q_m } \; ,
\end{align}
since $\avg{\delta q_{l,m}}=0$. Noting that this variance is equal to
$\cdiff k_BT$ and using the definition of the charge structure factor,
we obtain:
\begin{align}
S_{qq}(0) 
&= \frac{\avg{\Qtot^2}-\avg{\Qtot}^2}{N\avg{(\delta q)^2}}
= \frac{\cdiff k_B T}{{\revision{M}}\avg{(\delta q)^2}} \; .
\end{align}
This result holds for both electrodes (with the same $\cdiff$), 
even though $S_{qq}({\bf k})$ may differ for other wave-vectors if the adsorbed 
fluids adopt different structures. 
It is worth noting that this relation is similar to the one between the structure factor 
in a fluid of density $\rho$ and its compressibility $\chi_T$: 
$\lim_{k\to0}S(k)=\rho k_BT\chi_T$.

%
%


{\revision{
\subsection*{\large Comparison between the three systems}

Table~\ref{comparison} summarizes some properties of the three systems.
In particular, it compares the differential capacitance at $\Delta\Psi=0$~V
computed from the slope of $\left\langle Q\right\rangle$ vs. $\Delta\Psi$
and from the fluctuations of the total charge 
with the estimate $\epsilon_0\epsilon_r/d$ obtained by assuming that 
the electrolyte behaves as a pure dielectric with a permittivity equal to the bulk value.
For water with use the known value for the SPC/E water model; for the
MeCN based electrolyte we use the value for pure MeCN ($\epsilon_r=33$) which was computed in
Ref.~\cite{edwards1984a} for the present coarse-grained model. For the pure
ionic liquid we could not find in the literature a value for the model we use
and use instead the experimental value ($\epsilon_r=14$) of Ref.~\cite{singh_static_2008}.
}}

\vspace{0.5cm}
\begin{table*}[!ht]
\begin{center}
\begin{tabular}{|c|c|c|c|c|}
\hline
System & \ $\left.\frac{\partial\left\langle Q\right\rangle}{\partial\Delta\Psi}\right|_{\Delta\Psi=0}$ \ 
       & \ $\beta\left\langle\delta Q^2\right\rangle_{\Delta\Psi=0}$ \
       & $\epsilon_0\epsilon_r/d$  & $P(Q)$ \\
       & ($\mu$F.cm$^{-2}$) & ($\mu$F.cm$^{-2}$) & ($\mu$F.cm$^{-2}$)& \\
\hline
H$_2$O / Platinum             & 3.2 & 3.2 & 11.4 & Gaussian \\
\hline                                    
MeCN-[BMI][PF$_6$] / Graphite & 2.3 & 2.3 & 2.7  & Gaussian \\
\hline                                    
[BMI][PF$_6$] / Graphite      & --  & 2.1 & 1.1  & non-Gaussian \\
\hline
\end{tabular}
\vspace{-0.3cm}
\end{center}
\caption{
{\revision{Differential capacitance $\cdiff$ for $\Delta\Psi=0$~V 
computed from the slope of $\avg{Q}$ vs. $\Delta\Psi$ 
and from the fluctuation-dissipation, 
``naive'' estimate (see text) and behavior of the total charge
distribution $P(Q)$, for the three systems.}}}
\label{comparison}
\end{table*} 

{\revision{
For the solvent-based electrolytes, the values of $\cdiff$ obtained from the
slope of $\left\langle Q\right\rangle$ vs. $\Delta\Psi$
and from the fluctuations of the total charge are in excellent agreement, consistently
with the fluctuation-dissipation relation and with the results of 
Fig.~2 of the main text. For the pure ionic liquid, $\cdiff$ depends on 
the voltage (consistently with the non-Gaussian probability distribution), 
so that it would be necessary to perform simulations for many small
values of $\Delta\Psi$ in order to estimate
$\left.\frac{\partial\left\langle Q\right\rangle}{\partial\Delta\Psi}\right|_{\Delta\Psi=0}$.
On the contrary, the fluctuation-dissipation relation allows to compute $\cdiff$ from the
simulation at a single potential.

The ``naive'' estimate $\epsilon_0\epsilon_r/d$ 
is reasonable only in the case of the MeCN-based
electrolyte. In the water case, discussed in the main text, it is not possible
to neglect the molecular nature of the interface and the suppression of dipole
fluctuations on the surface. In the pure ionic liquid case, the description as a
pure dielectric is not relevant (not to mention interfacial effects). The
non-Gaussian distribution of the total charge in that case reflects correlations
between ions that are not present at low concentration (see the main text). 
}}





\end{document}